\documentclass[aps,twocolumn,nofootinbib]{revtex4}
\usepackage[latin1]{inputenc}
\usepackage{epsfig}
\newcommand{\beq}{\begin{equation}}
\newcommand{\eeq}{\end{equation}}
\newcommand{\la}{\langle}
\newcommand{\ra}{\rangle}

\begin{document}

\title{Entropy production in continuous systems with unidirectional transitions}
\author{Mário J. de Oliveira}
\affiliation{Universidade de São Paulo, Instituto de Física,
Rua do Matão, 1371, 05508-090 São Paulo, SP, Brazil}

\begin{abstract}

We derive the expression for the entropy production for
stochastic dynamics defined on a continuous space of states
containing unidirectional transitions. The expression is
derived by taking the continuous limit of a stochastic
dynamics on a discrete space of states and is based on an
expression for the entropy production appropriate for
unidirectional transition. Our results shows that the
entropy flux is the negative of the 
divergence of the vector firld whose components are the rates
at which a dynamic variable changes in time. For a
Hamiltonian dynamical system, it follows from this result
that the entropy flux vanish identically.

\end{abstract}

\maketitle

\section{Introduction}

Entropy and energy are fundamental concepts of thermodynamic.
Both entropy and energy share the property of being conserved
quantities of systems in thermodynamic equilibrium. Out of
thermodynamic equilibrium, energy remains a conserved quantity,
but entropy does not. However, entropy never decreases, a
property that expresses the second law. The increase of the
entropy $S$ of a system in time is therefore not only due the
flux of entropy but also due to the creation of entropy inside
the system, which is expressed by the relation
\beq
\frac{dS}{dt} = \Pi - \Psi,
\label{5}
\eeq
where $\Psi$ is the flux of entropy from the system 
to the outside and $\Pi$ is the rate at which entropy
is being created inside the system, and $\Pi\geq0$ which
is the expression of the second law.
The increase of the energy $U$ of the system is expressed by
\beq
\frac{dU}{dt} = \Phi,
\eeq
where $\Phi$ is the flux of energy from the outside to
the system, and reflects the conservation of energy.

Within stochastic thermodynamic
\cite{tome2015L,peliti2021,tome2006,tome2010,tome2015,tome2018}
entropy is defined through Gibbs
formula which is related to the probability distribution.
As this quantity varies in time so does the entropy from which
we determine $dS/dt$. The rate of entropy production $\Pi$, in
a discrete space of states, is usually determined by the
Schnakenberg formula \cite{schnakenberg1976}. If the space of
states is continuous, it can be determined by a formula which
can be understood as its extension to the continuum 
\cite{tome2010}.
As to the flux of entropy $\Psi$, it is obtained through
equation (\ref{5}), once $dS/dt$ and $\Pi$ are given.

The Schnakenberg formula has been extensively studied and applied
\cite{luo1984,mou1986,lebowitz1999,maes2003,crochik2005,%
seifert2005,zia2007,andrieux2007,harris2007,gaveau2009,%
esposito2009,szabo2010,hinrichsen2011,tome2012,esposito2012}.
It is appropriate for transitions that have their reverses 
but breaks down when this condition is not fulfilled, that is,
when the backward transition rate vanishes. The fluctuations
theorems are based on the probability of forward path and its
reversal also becomes impaired \cite{ohkubo2009,murashita2014}.
This problem has been addressed by several authors and
some proposals for its solution have been put forward
\cite{benavraham2011,zeraati2012,rahav2014,saha2016,
pal2017,busiello2020,pal2021,pal2021a,manzano2024}.
A simple solution of the problem is to conceive a formula 
\cite{tome2024} which is appropriate for unidirectional
transition, and indeed this has recently been proposed. Denoting by
$W_{x'x}$ the rate of the transition $x\to x'$ and by $P_x$
the probability distribution, then the rate of production of
entropy associated to this unidirectional transition is given
by \cite{tome2024}
\beq
\Pi_{x'x} =
W_{x'x}P_x\ln\frac{P_x}{P_{x'}} - W_{x'x}(P_x-P_{x'}),
\label{8}
\eeq
which is nonnegative because if we write $r=P_x/P_{x'}$
then this expression is proportional to $r\ln r - (r-1)\geq 0$,
for $r\geq 0$. The corresponding flux of entropy is \cite{tome2024}
\beq
\Psi_{x'x} = - W_{x'x}(P_x- P_{x'}).
\label{4}
\eeq

Our aim here is to extend formulas (\ref{8}) and (\ref{4}) to a
stochastic dynamics in a continuous space of states. To this end
we consider unidirectional transitions that can occur in several
directions each one of which occurring with a certain transition
rate. In the deterministic the dynamic equations reduce to
\beq
\frac{dx_i}{dt} = f_i(x),
\eeq
and the entropy flux $\Psi$ from the system to the outside
is found to be the negative of the divergence of the
vector field $f$ with components $f_i$,  
\beq
\Psi = - \sum_i \frac{\partial f_i}{\partial x_i}.
\label{25}
\eeq

The negative of the divergence of the vector field $f$ is
understood as the contraction of the volume of the state space
and has been suggested by Gallavotti and Cohen to be the rate
of entropy production \cite{gallavotti1995} and as such it
should be positive. However, they did not give a proof of the
positivity of this quantity but the positivity was shown by
Ruelle provided the system is in the steady state
\cite{ruelle1996,ruelle1997}. In accordance with the approach
that we follow here, the negative of divergence of the vector
field $f$ is identified as the entropy flux and {\it not} as
the rate of the entropy production. However, in the steady
state both expressions are equal to each other, and as the
production of entropy is positive so is the flux of entropy
given by (\ref{25}).

\section{Entropy production}

Let us consider a system described by a probability density
distribution $P(x)$ defined on a discrete vector space $x$
of dimension $n$, which varies in time according to the
master equation
\beq
\frac{dP_x}{dt} = \sum_{x'}(W_{xx'}P_{x'} - W_{x'x}P_x),
\eeq
where $W_{x'x}$ is the rate of the transition $x\to x'$.
From this equation one derive the time evolution of
the average 
\beq
\la F\ra = \sum_x F_x P_x
\eeq
of any state function $F_x$. Multiplying this equation
by $F_x$ and summing in $x$ we find
\beq
\frac{d}{dt}\la F\ra = \sum_{xx'} W_{x'x} P_x (F_{x'} - F_x).
\label{36}
\eeq

The entropy $S$ of the system is not the average of any state
function and is given by the formula
\beq
S = - \sum_x P_x\ln P_x,
\eeq
where we are omitting the Boltzmann constant. Gibbs used this
formula for systems in thermodynamic equilibrium
and called this expression the average of the index of
probability, $-\ln P_x$. In thermodynamic equilibrium
the probability distribution is the Gibbs distribution 
defined so that the index of probability is proportional
to the energy function. In this sense, the entropy is
indeed the average of a state function. However, as $P_x$
is not a state function, then generally speaking,
$\ln P_x$ is not a state function either and entropy is
not the average of a state function.

As $P_x$ depends on time so does the entropy $S$.
Its time derivative is given by
\beq
\frac{dS}{dt} = - \sum_x \frac{dP_x}{dt}\ln P_x.
\eeq
Using the master equation, it can be written as
\beq
\frac{dS}{dt} = \sum_{xx'}W_{x'x}P_x\ln \frac{P_x}{P_{x'}}.
\eeq

If the transition rates $W_{x'x}$ and $W_{xx'}$ are both
nonzero their contribution to the entropy production 
are determined by  
\beq
\Pi = \sum_{xx'}W_{x'x}P_x\ln \frac{W_{x'x}P_x}{W_{xx'}P_{x'}},
\eeq
which can be written as
\beq
\Pi = \frac12 \sum_{xx'}(W_{x'x}P_x - W_{xx'}P_{x'})
\ln \frac{W_{x'x}P_x}{W_{xx'}P_{x'}}.
\eeq
This expression was introduced by Schnakenberg
\cite{schnakenberg1976}, and is nonnegative because
each term of the sum is of the form $(a-b)\ln a/b\geq0$
The flux of entropy $\Psi$ from the system to the outside
is determined by $\Psi=\Pi-dS/dt$ and is given by
\beq
\Psi = \sum_{xx'} W_{x'x} P_x \ln \frac{W_{x'x}}{W_{xx'}}.
\label{35}
\eeq

If the transition rate $W_{x'x}$ is nonzero but its reverse
vanishes, that is, if the transition is unidirectional, then
the entropy production is determined by \cite{tome2024}
\beq
\Pi = \sum_{xx'}
W_{x'x}P_x\ln\frac{P_x}{P_{x'}} - W_{x'x}(P_x-P_{x'}),
\eeq
which is nonnegative because each term of the summation
is of the type $r\ln r - (r-1)\geq 0$.
The corresponding flux of entropy is \cite{tome2024}
\beq
\Psi = - \sum_{xx'}W_{x'x}(P_x- P_{x'}).
\label{37}
\eeq

\section{Energy function}

Within ordinary mechanics, the construction of the energy
function relies on the conservative forces through which we
define the potential energy which in turn is added to the
kinetic energy. Within stochastic dynamics the construction
of the energy function relies on the transition rates.
In systems that reaches thermodynamic equilibrium, the
relation between these two quantities, energy function and 
transition rates, comes from the assumption of detailed
balance condition also called microscopic reversibility.
Conservative forces and detailed balance condition are
analogous concepts. The work of a conservative force between
two points is independent of the path connecting them.
The probabilities of two trajectories connecting two states
are the same if detailed balance is satisfied.

In the case of nonequilibrium stochastic dynamics we cannot
use detailed balance as it does not hold. In this case 
we may use the following relation between the transition
rates and the energy function $E_x$,
\beq
\ln \frac{W_{x'x}}{W_{xx'}} = -\beta_{xx'} (E_{x'} - E_x).
\label{33}
\eeq
where $\beta_{x'x}=\beta_{xx'}$. We remark that this is not
the condition of detailed balance unless $\beta_{xx'}$ is the
same for all pairs of states $x$ and $x'$. Relation (\ref{33})
is usually called local detailed balance, but we avoid this
terminology.

Equation (\ref{33}) leads us to a relation between entropy flux
and energy flux. Before presenting this relation, we need to
define the flux of energy. If in equation (\ref{36}),
we replace $E$ by $F$ we find the time evolution of $U=\la E\ra$,
$dU/dt=\Phi$, where $\Phi$ is understood as the flux of energy into
the system, given by
\beq
\Phi = \sum_{x'x}  W_{x'x} P_x (E_{x'} - E_x).
\eeq
Replacing (\ref{33}) in (\ref{35}) we get 
\beq
\Psi = - \sum_{xx'}\beta_{xx'}  W_{x'x} P_x(E_{x'} - E_x),
\label{17}
\eeq
which relates the entropy flux and the energy function.
In the case of detailed balance, when all $\beta_{xx'}$
are equal to each other, this expression lead us to
the relation $\Psi=-\beta \Phi$ between $\Psi$ and
the total flux of energy $\Phi$. This relation is
equivalent to the Clausius relation between 
entropy and heat valid for equilibrium systems.

For unidirectional transitions, the relation (\ref{33})
cannot be used because either $W_{x'x}$ or $W_{xx'}$
vanish. In this case we propose the following relation
between transition rates and energy function
\beq
\sum_{x'} (W_{x'x}-W_{xx'})
= \sum_{x'}\beta_{x'x}W_{x'x}(E_{x'} - E_x).
\label{34a}
\eeq 
Multiplying both sides of this equation by $-P_x$
and summing in $x$ we find
\beq
-\sum_{xx'} (W_{x'x}-W_{xx'})P_x
= -\sum_{x'x}\beta_{x'x}W_{x'x}P_x(E_{x'} - E_x).
\eeq 
Comparing the left hand side of this equation with (\ref{37})
we see that it equals $\Psi$, and we reach again the expression
(\ref{17}) for $\Psi$.

\section{Unidirectional stochastic motion}

\subsection{Master equation}

We consider here a stochastic motion in a continuous space
of states which is a vector space of dimension $n$. A vector
of this space is denoted by $x$ and its components by $x_i$,
\beq
x = (x_1,x_2,\ldots,x_n).
\eeq
The stochastic motion consists only of unidirectional
transitions. However, from a given state there may arise
several unidirectional motions each one occurring with a
certain probability. As the vector space has dimension
$n$, there are $n$ directions. Each direction is represented
by a unit vector $c^\nu$, with components $c_i^\nu$,
\beq
c^\nu = (c_1^\nu,c_2^\nu,\ldots,c_n^\nu),
\eeq
where $\nu=1,2,\ldots,n$. These
vectors do not depend on $x$ and are chosen
to form an orthogonal set, $c^\nu\cdot c^\mu=\delta_{\nu\mu}$.

Given a state $x$, the possible transitions are those
to a state $x'$ that differs from $x$ by a distance
$\varepsilon$, that is, $|x'-x|=\varepsilon$, and such
that $x'-x$ is in the direction of one of the vectors
$c^\nu$. The rate of the transition
$x\to x'= x + \varepsilon c^\nu$ is denoted by $w_\nu(x)$
and depend on $x$. from which follows the master equation

\beq
\frac{dP(x)}{dt} = \frac1\varepsilon \sum_\nu
[w_\nu(x-\varepsilon c^\nu) P(x-\varepsilon c^\nu) - w_\nu(x)P(x)].
\label{3a}
\eeq
Defining 
\beq
f_i = \sum_\nu c_i^\nu  w_\nu.
\label{23a}
\eeq
the time evolution of the average $s_i=\la x_i\ra$ is obtained
from the master equation and is given by
\beq
\frac{ds_i}{dt} = \la f_i\ra.
\eeq

The entropy is defined by 
\beq
S = - \sum_x P(x)\ln P(x),
\eeq
and its time evolution is given by
\beq
\frac{dS}{dt} = - \sum_x \frac{dP(x)}{dt}\ln P(x).
\eeq
Replacing (\ref{3a}) in this equation, we find
\beq
\frac{dS}{dt} = - \frac1\varepsilon\sum_{\nu x} 
[w_\nu(x-\varepsilon c^\nu) P(x-\varepsilon c^\nu)
- w_\nu(x)P(x)]\ln P(x),
\label{15a}
\eeq
which can be written as
\beq
\frac{dS}{dt} = - \frac1\varepsilon\sum_{\nu x} 
w_\nu(x) P(x) \ln \frac{P(x+\varepsilon c^\nu)}{P(x)}.
\label{15b}
\eeq

The rate of entropy production $\Pi$ is given by the expression
that we have proposed for unidirectional transitions \cite{tome2024}
\beq
\Pi = \frac1\varepsilon\sum_{\nu x} w_\nu(x)
[P(x) \ln \frac{P(x)}{P(x+\varepsilon c^\nu)}
- P(x) + P(x+\varepsilon c^\nu)],
\label{7}
\eeq
and the entropy flux $\Psi$ is obtained throug
$\Psi=\Pi-dS/dt$ and is
\beq
\Psi = - \frac1\varepsilon\sum_\nu\sum_x w_\nu(x)
[P(x) - P(x+\varepsilon c^\nu)],
\eeq
which can be written as the average
\beq
\Psi = - \frac1\varepsilon \sum_\nu\la w_\nu(x)
- w_\nu(x-\varepsilon c^\nu)\ra.
\label{12}
\eeq

\subsection{Fokker-Planck equation}

Next we derive from the above equations the expressions for
small values of $\varepsilon$. Expanding the expression between
curl brackets in equation (\ref{3a}) up to second order in
$\varepsilon$, then the master equation (\ref{3a}) becomes
the Fokker-Planck equation
\beq
\frac{dP}{dt} = - \sum_i \sum_\nu
\frac{\partial c^\nu_i w_\nu P}{\partial x_i}
+\frac\varepsilon2 \sum_{ij} \sum_\nu\frac{\partial^{\,2}
c_i^\nu c_j^\nu w_\nu P}{\partial x_i\partial x_j}.
\eeq
Recalling the definition of $f_i$ given by (\ref{23a}),
then the Fokker-Planck can be written as
\beq
\frac{dP}{dt} = - \sum_i \frac{\partial f_i P}{\partial x_i}
+\frac\varepsilon2 \sum_{ij} \frac{\partial^{\,2}\Gamma_{ij} P}
{\partial x_i\partial x_j},
\label{27}
\eeq
where
\beq
\Gamma_{ij} = \sum_\nu c_i^\nu c_j^\nu w_\nu.
\label{23b}
\eeq

The vectors $c^\nu$ are the eigenvectors
and $w_\nu$ are the eigenvalues of the matrix $\Gamma$.
To show this result it suffices to write
\beq
\sum_j \Gamma_{ij} c_j^\mu
=  \sum_\nu c_i^\nu w_\nu \sum_ j (c_j^\nu c_j^\mu) 
= \sum_\nu c_i^\nu w_\nu \delta_{\nu\mu} = w_\mu c_i^\mu,
\eeq
where we have used the orthogonality of the unit vectors
$c^\nu$. Considering that $w_\nu\geq0$, then the matrix
$\Gamma$ is positive semi definite.

The expression (\ref{15b}) for $dS/dt$ becomes
\beq
\frac{dS}{dt} = \sum_i\int(\frac{\partial f_i P}{\partial x_i}
- \frac\varepsilon2 \sum_j\frac{\partial^2 \Gamma_{ij} P}
{\partial x_i\partial x_j})\ln Pdx,
\eeq
An integration by parts gives
\beq
\frac{dS}{dt} = - \sum_i\int\frac1P(f_i P
- \frac\varepsilon2 \sum_j\frac{\partial \Gamma_{ij} P}
{\partial x_j}) \frac{\partial P}{\partial x_i}dx.
\label{32}
\eeq

The entropy production $\Pi$, given by (\ref{7}) becomes
\beq
\Pi = \frac{\varepsilon}2 \sum_{ij}\int \frac{\Gamma_{ij}}{P}
\frac{\partial P}{\partial x_i} 
\frac{\partial P}{\partial x_j}dx.
\label{39}
\eeq
which is clearly nonnegative because $\Gamma$ is positive
semi definite, and the entropy flux $\Psi$, given by
(\ref{12}), becomes the average
\beq
\Psi = -\sum_i \la\frac{\partial f_i}{\partial x_i}\ra
+ \frac\varepsilon2 \sum_{ij}\la\frac{\partial^2 \Gamma_{ij}}
{\partial x_i\partial x_j}\ra.
\label{31}
\eeq

It is easily checked that the expressions above 
fulfills the relation $dS/dt=\Pi-\Phi$. Indeed,
writing equation (\ref{32}) as
\beq
\frac{dS}{dt} = - \sum_i\int\frac1P(f_i P
- \frac\varepsilon2 \sum_j \frac{\partial \Gamma_{ij}}
{\partial x_j} P - \frac\varepsilon2 \sum_j \Gamma_{ij}
\frac{\partial P}{\partial x_j}) \frac{\partial P}{\partial x_i}dx,
\eeq
we see that the first two term on the right-hand side of this
equation gives $\Psi$ and the last one gives $\Pi$.

\subsection{Deterministic limit}

Next we analyse the solution of the Fokker in the regime of small
$\varepsilon$. As the parameter $\varepsilon$ is a measure of the
fluctuations, we expect that for small values of $\varepsilon$
the probability distribution $P(x)$ be very peaked at the average
of $x_i$. If we denote by $s_i$ the average of $x_i$ then in the
limit $\varepsilon\to0$, we expect that $s_i$ varies in time
according to
\beq
\frac{ds_i}{dt} = f_i(s),
\label{10}
\eeq
where
\beq
f_i(s) = \sum_\nu c_i^\nu w_\nu(s).
\label{10a}
\eeq
The fluctuations of $x_i$ around $s_i$ are expected
to be proportional to $\sqrt{\varepsilon}$. These 
considerations suggest us to introduce the following
transformation of variables from $x_i$ to $y_i$
\beq
y_i = \frac{x_i-s_i}{\sqrt\varepsilon},
\label{9}
\eeq
where $s_i(t)$ depends on time and is the solution of
equation (\ref{10}).

The probability distribution of the new variable $y$
is denoted by $\rho(y)$ and is related to $P(x)$
by $\rho(y)dy=P(x)dx$ or
\beq
\rho(y) = \varepsilon^{n/2} P(s+\sqrt\varepsilon y).
\eeq
From this relation we find
\beq
\frac{\partial \rho}{\partial t}
= \varepsilon^{n/2}\frac{\partial P}{\partial t}
+ \varepsilon^{n/2}\sum_i \bar{f}_i \frac{\partial P}{\partial x_i},
\eeq
where the bar over $f$ indicates that it should be understood
as a function of $s$ and not of $x$, that is, $\bar{f}_i=f_i(s)$.
Replacing in this equation $\partial P/\partial t$ given by the
Fokker-Planck equation (\ref{27}), we reach the following
equation for $\rho$
\beq
\frac{\partial \rho}{\partial t} =
\frac1{\sqrt\varepsilon}\sum_i \bar{f}_i\frac{\partial \rho}
{\partial y_i} - \frac1{\sqrt\varepsilon}\sum_i
\frac{\partial f_i \rho}{\partial y_i} + \frac12 \sum_{ij}
\frac{\partial^2 \Gamma_{ij} \rho}{\partial y_i\partial y_j}.
\eeq
In this form the only quantities that depend on $\varepsilon$
are $f_i$ and $\Gamma_{ij}$ because they are functions of
$(s+\varepsilon y)$. 
The limit $\varepsilon\to 0$ is obtained by observing that
\beq
\frac{f_i(s+\varepsilon y)-f_i(s)}{\sqrt\varepsilon}
\to \sum_j f_{ij}(s)y_i,
\eeq
where $f_{ij}(s)=\partial f_i(s)/\partial s_i$, and that
\beq
\Gamma_{ij}(s+\varepsilon y) \to \Gamma_{ij}(s).
\eeq
The equation for $\rho$ becomes
\beq
\frac{\partial \rho}{\partial t} =
- \sum_{ij} \bar{f}_{ij}  \frac{\partial y_j \rho}{\partial y_i}
+ \frac12 \sum_{ij}\bar{\Gamma}_{ij}
\frac{\partial^2  \rho}{\partial y_i\partial y_j},
\label{11}
\eeq
where again the bars over $f_{ij}$ and $\Gamma_{ij}$
indicates that they are functions of $s$, and thus
depend on $t$ through $s(t)$.

The solution of equation (\ref{11}) is a 
multivariate Gaussian distribution of the form
\beq
\rho(y) = \frac1Z
\exp\{-\frac12\sum_{ij}(\chi^{-1})_{ij} \, y_i y_j\},
\label{38a}
\eeq
where
\beq
Z = \int \exp\{-\frac12\sum_{ij}(\chi^{-1})_{ij} \, y_i y_j\} dy,
\label{15}
\eeq
and the covariances $\chi_{ij}=\la y_i y_j\ra$ depend on $t$.
Performing the integral in (\ref{15}) we obtain the result
\beq
Z= {(2\pi)^{n/2} [{\rm Det}(\chi)]^{1/2}}.
\eeq
Thus the solution of the Fokker-Planck is fully determined
if $\chi_{ij}$ is found as a function of $t$. An
equation that determines the covariance is obtained
from equation (\ref{11}). After multiplying (\ref{11})
by $y_iy_j$ and integrating in $y$, we find
\beq
\frac{d\chi_{ij}}{dt}=
\sum_k \bar{f}_{ik} \chi_{jk } + \sum_k \bar{f}_{jk} \chi_{ik}
+ \bar{\Gamma}_{ij},
\label{16}
\eeq
where appropriate integrations by parts have been performed.

We remark that $s_i$, which was introduced as the solution of
equation (\ref{10}), is identified as the average $\la x_i\ra$.
Indeed, from (\ref{9}) it follows that
$\la x_i\ra = s_i +\sqrt\varepsilon\la y_i\ra$.
But from the distribution $\rho$, $\la y_i\ra=0$

Let us determine $\Psi$ and $\Pi$ in the limit $\varepsilon\to0$.
From the Gaussian distribution we find in this limit
$\la f_{ij}(x)\ra \to f_{ij}(s)$ and that the second
term on the right-hand side of (\ref{31}) vanishes, and
\beq
\Psi = -\sum_i \frac{\partial \bar{f}_i}{\partial s_i}.
\label{47}
\eeq
This is the main result of the present approach. 
It says that the entropy flux is the negative of the
divergence of the vector field $f$.

The expression (\ref{39}) for the rate of
entropy production is written in terms of $\rho$ as
\beq
\Pi = \frac12 \sum_{ij} 
\int \rho \frac{\partial \ln\rho}{\partial y_i}
\bar{\Gamma}_{ij}\frac{\partial\ln\rho}{\partial y_j}dy.
\eeq
Using 
\beq
\ln \rho = -\frac12\sum_{kl}(\chi^{-1})_{kl} \, y_k y_l - \ln Z,
\label{14}
\eeq
we reach the following expression
\beq
\Pi
= \frac12 \sum_{ij} \bar{\Gamma}_{ij} (\chi^{-1})_{ji}.
\label{13}
\eeq

Let us write the equation (\ref{16}) for the time evolution of
$\chi_{ij}$ in the matrix form
\beq
\frac{d}{dt} \chi = F\chi + \chi F^{\sf T} + \bar\Gamma,
\eeq
where $F$ is the matrix with elements
$F_{ij}=\bar{f}_{ij}=\partial\bar{f}_i/\partial s_j$.
Multiplying this equation on the left by $\chi^{-1}$
and taking the trace, we reach the relation
\beq
\frac12 {\rm Tr}(\chi^{-1} \frac{d}{dt} \chi) = 
\frac12 {\rm Tr}(\bar\Gamma \chi^{-1}) + {\rm Tr} F.
\label{48}
\eeq
From (\ref{47}), the entropy flux $\Psi$ is
\beq
\Psi = - \sum_i F_{ii} =  - {\rm Tr} F,
\eeq
and from (\ref{13}), the rate of entropy $\Pi$ is
\beq
\Pi = \frac12 {\rm Tr} (\Gamma \chi^{-1}),
\eeq
and we see that the right-hand side of (\ref{48}) 
is $\Pi-\Psi$ from which follows that the left
hand-side of this equation is $dS/dt$,
\beq
\frac{dS}{dt} = \frac12 {\rm Tr}\chi^{-1} \frac{d}{dt} \chi.
\label{49}
\eeq

The expression (\ref{49}) can be obtained directly
from the  Gaussian distribution (\ref{38a}) as follows. 
The entropy 
\beq
S = - \int P \ln P dx,
\eeq
written in terms of $\rho$ is
\beq
S = - \int \rho \ln (\rho \varepsilon^{-n/2}) dy.
\eeq
Using (\ref{14}),
\beq
S =  \frac{n}2 (\ln \varepsilon + 1) + \ln Z.
\label{45}
\eeq
Deriving $S$ with respect to time,
\beq
\frac{dS}{dt} = \frac1Z\frac{dZ}{dt}
=- \frac12\sum_{ij} \frac{d(\chi^{-1})_{ij}}{dt}\chi_{ij}
= \frac12\sum_{ij} (\chi^{-1})_{ij} \frac{d\chi_{ij}}{dt},
\eeq
where we used the definition of $Z$ given by (\ref{15}).
The last expression is identical to the expression in the
right-hand side of (\ref{49}).

\subsection{Energy function}

To associate an energy function $E(x)$ to the stochastic dynamics
described by the master equation (\ref{3a}) we first determine the
energy flux. This quantity is obtained by writing the time
evolution of $U=\la E\ra$. From the master equation we obtain
$dU/dt = \Phi$, where $\Phi$ is the energy flux,
\beq
\Phi = \frac1\varepsilon \sum_\nu \la
[E(x+\varepsilon c^\nu) - E(x)] w_\nu(x)\ra.
\eeq
The equation (\ref{34a}) that relates the transition rates
and the energy function, in the present case reads
\beq
\sum_\nu [w_\nu(x) - w_\nu(x-\varepsilon c^\nu)] =
\sum_\nu \beta_\nu(x+\varepsilon c^\nu,x) w_\nu(x)
[E(x+\varepsilon c^\nu) - E(x)].
\eeq

The expressions of the above results for small values of
$\varepsilon$ are
\beq
\Phi = \sum_i \la f_i \frac{\partial E}{\partial x_i}\ra.
\eeq
and
\beq
\sum_i \sum_\nu c_i^\nu \frac{\partial w_\nu}{\partial x_i}
= \sum_i \sum_\nu \beta_i^\nu c_i^\nu w_\nu
\frac{\partial E}{\partial x_i}.
\eeq
and a sufficient condition for this last equation to be fulfilled is
\beq
\frac{\partial w_\nu}{\partial x_i}
= \beta_i^\nu w_\nu \frac{\partial E}{\partial x_i}.
\eeq
In the limit $\varepsilon\to0$, $\la E(x)\ra\to E(s)$, 
and the two equations above become
\beq
\frac{d\bar{E}}{dt} = \Phi, \qquad
\Phi = \sum_i \bar{f}_i \frac{\partial \bar{E}}{\partial s_i}
\label{29}
\eeq
and
\beq
\frac{\partial \bar{w}_\nu}{\partial s_i}
= \bar{\beta}_i^\nu\bar{w}_\nu \frac{\partial \bar{E}}{\partial s_i},
\label{30}
\eeq
where as before the bars indicate functions of $s$,
that is, $\bar{f}_i=f_i(s)$ and $\bar{E}=E(s)$
and we recall that $s(t)$ depends on time and is the
solution of $ds/dt=f(s)$. We also recall that 
the entropy flux $\Psi$ is given by (\ref{47}) and is
\beq
\Psi = -\sum_i \frac{\partial \bar{f}_i}{\partial s_i}
= -\sum_i \sum_\nu c_i^\nu \frac{\partial \bar{w}_\nu}{\partial s_i}
\eeq
Replacing (\ref{30}) in this expression,
\beq
\Psi = - \sum_\nu \sum_i \bar{\beta}_i^\nu  c_i^\nu
\bar{w}^\nu \frac{\partial \bar{E}}{\partial s_i}.
\label{40}
\eeq

It is worth writing $\Phi$ and $\Psi$ as a sum of terms
\beq
\Phi = \sum_i \sum_\nu \Phi_i^\nu,
\qquad \Psi = \sum_i \sum_\nu \Psi_i^\nu
\eeq
where 
\beq
\Phi_i^\nu = \bar{c}_i^\nu \bar{w}_\nu
\frac{\partial \bar{E}}{\partial s_i}
\eeq
which is understood as the flux of energy associated
to the change of $x_i$ in the direction $c^\nu$,
and 
\beq
\Psi_i^\nu = -\beta_i^\nu \Phi_i^\nu
\eeq
which is understood as the flux of entropy associated to this
change.

If $\Psi$ vanishes, that is, if the vector $f$ is such that its
divergence vanish, then we may choose $\beta_i^\nu$ independent
of $i$ and $\nu$, and equation (\ref{40}) becomes
\beq
\sum_i \bar{f}_i \frac{\partial \bar{E}}{\partial s_i} =0.
\eeq
Comparing with equation (\ref{29}), the flux of energy $\Phi$
vanishes and the energy $\bar{E}(s)$ is a constant of the motion.
This equation also shows that the vector representing the gradient
of $\bar{E}$ is perpendicular to $\bar{f}$.

\section{Chemical kinetics}

The present framework can be applied to the theory of chemical
kinetics for the case where the reactions are unidirectional.
We consider a vessel containing molecules of several chemical
species that react among themselves. The dynamic variable $x_i$
is now understood as the number of molecules of species $i$.
The transition $x'\to x+\varepsilon c^\nu$ which
occurs with transition rate $w_\nu$ is interpreted as a 
unidirectional chemical reaction occurring with rate $w_\nu$
in which the number of molecules $x_i$ changes by an amount
$\varepsilon c_i^\nu$. Therefore, the constants $c_i^\nu$ are
interpreted as proportional to the difference between the
stoichiometric coefficients of the products and the reactants
of the reaction $\nu$. The equations (\ref{10}) and (\ref{10a})
are rewritten as
\beq
\frac{dx_i}{dt} = f_i, \qquad f_i = \sum_\nu c_i^\nu w_\nu(x),
\eeq
and are understood as the equations of the chemical kinetics.
We are using $x_i$ in the place of $s_i$ and we will do that
from now on.

To proceed in our analysis, we assume the transition rates as
given by the law of mass action, that is,
\beq
w_\nu = k_\nu\prod_i x_i^{\alpha_i^\nu},
\label{42}
\eeq
where $k_\nu$ is the rate constant of the reaction $\nu$,
and $\alpha_i^\nu$ are the stoichiometric coefficients of the
reactants only. 

The entropy flux $\Psi$ given by the equation (\ref{47})
becomes 
\beq
\Psi = -\sum_i \frac{\partial f_i}{\partial x_i}
= - \sum_i \sum_\nu c_i^\nu   \frac{\alpha_i^\nu}{x_i} w_\nu.
\eeq
and the variation of the energy $E(x)$ with time is
\beq
\frac{dE}{dt} = \Phi,
\eeq
where the flux of energy $\Phi$ is given by (\ref{29})
and is
\beq
\Phi = \sum_i \sum_\nu c_i^\nu w_\nu \frac{\partial E}{\partial x_i}.
\eeq

To determine the form of the energy function $E(x)$ we have to solve
equation (\ref{30}) which for the transition rate (\ref{42}) becomes
\beq
\frac{\alpha_i^\nu}{x_i}
= \beta_i^\nu \frac{\partial E}{\partial x_i},
\eeq
whenever $w_\nu$ is nonzero, whose solution is
\beq
E = \sum_i h_i \ln x_i,
\eeq
and
\beq
\beta_i^\nu = \frac{\alpha_i^\nu}{h_i} 
\eeq
valid if $\alpha_i^\nu$ is nonzero. If $\alpha_i^\nu=0$,
which means that $w_\nu(x)$ does not depend on $x_i$, then
$\beta_i^\nu=0$. 

The expressions for the flux of energy and the flux
of entropy become
\beq
\Phi = \sum_i \sum_\nu c_i^\nu h_i \frac{w_\nu }{x_i},
\eeq
\beq
\Psi = - \sum_i \sum_\nu \beta_i^\nu  c_i^\nu  h_i \frac{w_\nu}{x_i} .
\eeq
In the stationary state, the energy flux vanishes, $\Phi=0$,
but the entropy flux $\Psi$ does not, unless all $\beta_i^\nu$
are equal.

\section{Discussion and conclusion}

We have addressed here the problem of determining the entropy
production for system containing unidirectional transitions
in continuous stochastic dynamics. The problem was solved by
using a formula that was previously introduced  for the
entropy production in discrete stochastic dynamics containing
unidirectional transitions. The formulas we derived contained
a small parameter $\varepsilon$ that measures the fluctuations
of the continuous variables. In the limit $\varepsilon\to0$
we obtained results that is understood to be valid for
deterministic motion. The main result is the expression for
the entropy flux, given by equation (\ref{47}) which says that
this quantity is the negative of the divergence of the
vector field $f$. 

As the expression (\ref{47}) was obtained by considering a
stochastic dynamics and then taking the deterministic
limit $\varepsilon\to0$, a question then arises whether it
is valid for a dynamic system given by the set of equations 
\beq
\frac{dx_i}{dt} = f_i(x),
\label{44}
\eeq
where no mention to stochastic motion is given. Suppose
that $f$, the vector with components $f_i$, can be
written as a sum of orthogonal vectors $f^\nu$,
\beq
f = \sum_\nu f^\nu.
\eeq
Defining $e^\nu$ as the unit vector in the direction of
$f^\nu$ then $f^\nu=e^\nu |f^\nu|$ and 
\beq
f_i = \sum_\nu e_i^\nu |f^\nu|,
\eeq
where $e_i^\nu$ are the components of $e^\nu$.
We see that this expression has the form of (\ref{23a})
and we may identify $|f^\nu|$ as a transition rate
and formulate a stochastic motion. 
Therefore, the entropy flux given by (\ref{47})
can be used in relation to the dynamic system
(\ref{44}) as long as the splitting of $f$ just
mentioned can be carried out. The same can be said
concerning the rate of entropy production, given
by (\ref{13}), and the entropy given by (\ref{45}).

For a Hamiltonian motion, the state space is the phase
space consisting of the coordinates $q_i$ and momenta
$p_i$. The equations of motion are
\beq
\frac{dq_i}{dt} = \frac{\partial H}{\partial p_i}, \qquad
\frac{dp_i}{dt} = - \frac{\partial H}{\partial q_i}
\eeq
where $H$ is the Hamiltonian function.
In this case, the entropy flux (\ref{47}) for
the Hamiltonian motion is
\beq
\Psi = - \sum_i (\frac{\partial}{\partial q_i}
\frac{\partial H}{\partial p_i} - \frac{\partial}{\partial p_i}
\frac{\partial H}{\partial q_i})
\eeq
which vanishes identically, $\Psi=0$.

We remark finally that the negative of the divergence of the
vector field $f$ was suggested by Gallavotti and Cohen to be
the rate of entropy production and thus not the entropy flux
as we did here. However, these two quantities become equal
when the system reaches a stationary state, when the entropy
becomes independent of time. In this case the entropy flux becomes
positive because the production of entropy is positive, a
result which is consistent with the Ruelle demonstration that the
Gallavotti and Cohen entropy production is positive in the
stationary state.


\end{document}